\documentclass[article,aps,nofootinbib,twocolumn,superscriptaddress]{revtex4-1}
\input epsf
\usepackage{graphics}
\usepackage{amsmath,amssymb}
\usepackage{color}
\usepackage{dcolumn}
\usepackage{hyphenat}
\usepackage{multirow}

\usepackage{graphicx}

\def\be{\begin{equation}}
\def\ee{\end{equation}}
\def\ba{\begin{eqnarray}}
\def\ea{\end{eqnarray}}

\newcommand{\beqa}{\begin{eqnarray}}
\newcommand{\eeqa}{\end{eqnarray}}

\newcommand{\beq}{\begin{equation}}
\newcommand{\eeq}{\end{equation}}

\newlength{\tskip}
\newlength{\colwidth}

\def\kjm#1#2#3{k^{(#1)}_{(#2)#3}}
\def\kfd#1{k_{F}^{(#1)}}
\def\kafd#1{k_{AF}^{(#1)}}
\def\kVdjm#1#2{\kjm{#1}{V}{#2}}

\begin{document}

\title{Future CMB constraints on cosmic birefringence and implications for fundamental physics}

\author{Levon Pogosian} \affiliation{Department of Physics, Simon Fraser University, Burnaby, BC, V5A 1S6, Canada}
\author{Meir Shimon} \affiliation{School of Physics and Astronomy, Tel Aviv University, Tel Aviv 69978, Israel}
\author{Matthew Mewes} \affiliation{Department of Physics, California Polytechnic State University, San Luis Obispo, CA 93407, USA}
\author{Brian Keating} \affiliation{Department of Physics, University of California, San Diego, CA 92093-0424, USA}

\begin{abstract}
The primary scientific target of the ground and space based Cosmic Microwave Background (CMB) polarization experiments currently being built and proposed is the detection of primordial tensor perturbations. As a byproduct, these instruments will significantly improve constraints on cosmic birefringence, or the rotation of the CMB polarization plane. If convincingly detected, cosmic birefringence would be a dramatic manifestation of physics beyond the standard models of particle physics and cosmology. We forecast the bounds on the cosmic polarization rotation (CPR) from the upcoming ground-based Simons Observatory (SO) and the space-based LiteBIRD experiments, as well as a ``fourth generation'' ground-based CMB experiment like CMB-S4 and the mid-cost space mission PICO.  We examine the detectability of both a stochastic anisotropic rotation field, as well as an isotropic rotation by a constant angle. CPR induces new correlations of CMB observables, including spectra of parity-odd type in the case of isotropic CPR, and mode-coupling correlations in the anisotropic rotation case. We find that LiteBIRD and SO will reduce the 1$\sigma$ bound on the isotropic CPR from the current value of 30 arcmin to 1.5 and 0.6 arcmin, respectively, while a CMB-S4-like experiment and PICO will reduce it to $\sim 0.1$ arcmin. The bounds on the amplitude of a scale-invariant CPR spectrum will be reduced by one, two  and three orders of magnitude by LiteBIRD, SO and CMB-S4-like/PICO, respectively.  We discuss potential implications for fundamental physics by interpreting the forecasted bounds on CPR in terms of the corresponding constraints on pseudoscalar fields coupled to electromagnetism, primordial magnetic fields (PMF), and violations of Lorentz invariance.  We find that CMB-S4-like and PICO can reduce the $1\sigma$ bound on the amplitude of the scale-invariant PMF from 1 nG to 0.1 nG, while also probing the magnetic field of the Milky Way. The upcoming experiments will also tighten bounds on the axion-photon coupling, with SO improving the bound from $f_a \gtrsim 50 H_I$ at present, where $H_I$ is the energy scale of inflation, to $f_a \gtrsim 500 H_I$, and CMB-S4-like and PICO raising it to $f_a \gtrsim {\rm few} \times 10^{3} H_I$, placing stringent constraints on the string theory axions.
\end{abstract}

\maketitle

\section{Introduction}

The impact of the Cosmic Microwave Background (CMB) on our knowledge of the primordial universe has been astounding. In the past quarter of a century, progress has hardly abated. Recent years have witnessed the discovery of temperature anisotropy by COBE \cite{Smoot:1992td,Bennett:1996ce}, then the first detection of CMB polarization by DASI \cite{Kovac:2002fg} and high resolution full sky CMB maps from WMAP \cite{Bennett:2003bz,Bennett:2012zja},  culminating in comprehensive measurements of temperature and E-mode polarization by Planck \cite{Adam:2015rua}. The primary focus of current CMB research is the measurements of the so-called B-modes -- the parity odd polarization patterns \cite{Kamionkowski:1996ks,Seljak:1996gy} that could be created by inflationary gravitational waves (GW) \cite{Crittenden:1993ni,Seljak:1996gy,Kamionkowski:1996zd} as well as a harbinger of potentially new  physics \cite{Seljak:1997ii,Avgoustidis:2011ax,Moss:2014cra,Lizarraga:2016onn,Amendola:2014wma,Raveri:2014eea}. On $\sim 10'$ angular scales, or $\ell \sim 1000$, gravitational lensing by large scale structures generates B-modes \cite{Zaldarriaga:1998ar} that were measured by POLARBEAR \cite{Ade:2014afa} and SPTPol \cite{Keisler:2015hfa} five years ago. The first measurements of B-modes on larger scales, $\ell \sim 100$, where the inflationary GW are expected to contribute the most, were made by BICEP2/Keck \cite{Ade:2014xna,Ade:2015tva}. However,  foregrounds, such as  polarized dust in our galaxy, are not yet characterized to an accuracy needed to unveil the primordial signal possibly hiding behind.

While  foregrounds pose a serious challenge, many experiments are rising to meet it. For example the Simons Observatory \cite{Ade:2018sbj}, currently under construction in the Chilean Atacama desert, has a predicted sensitivity of $\sigma_r=0.003$ to the tensor-to-scalar ratio $r$ characterizing the amplitude of gravitational wave (GW) B-modes, which would improve current bounds \cite{Array:2015xqh} by more than an order of magnitude. As the inflationary paradigm is perfectly consistent with $r$ being below the observable range \cite{Knox:2002pe}, it is plausible that no GW contribution will ever be seen with B-modes. Fortunately, other fundamental physics will be constrained by improved B-mode measurements (see, {\it e.g.}, \cite{Abazajian:2016yjj} for a review), such as the number of relativistic particle species in the early universe, the sum of the neutrino masses, annihilation rates of dark matter candidates, and possible modifications of gravity. Another effect, which is the subject of this paper, is cosmic birefringence, or cosmic polarization rotation (CPR), that can be caused by parity violating extensions of the standard model \cite{Harari:1992ea,Carroll:1989vb,Carroll:1998zi,Pospelov:2008gg} or primordial magnetic fields \cite{Kosowsky:1996yc}.

Unlike the inflationary GW B-modes where the target signal is, at best, a few percent of the foreground contribution, most of the signal required for CPR measurements comes from smaller scales that are less affected by  galactic foregrounds. Still, foregrounds and instrumental systematic effects, such as the beam asymmetry and imperfect scanning of the sky, play an important role.

CPR is manifested in different types of correlations of CMB observables, depending on whether the rotation angle is uniform or varies across the sky. In either case, the rotation converts some of the E-modes into B-modes, generating a contribution to the B-mode power spectrum. A similar B-to-E conversion also takes place but is negligible, as the primordial B-modes are constrained to be subdominant. A uniform rotation angle leads to parity-odd spectra of EB and TB type. An anisotropic rotation, on the other hand, introduces mode-coupling that leads, in particular, to non-trivial 4-point correlations \cite{Kamionkowski:2008fp,Yadav:2009eb,Gluscevic:2009mm,Gluscevic:2012me}. A detection of CPR would signal new physics beyond the standard models of cosmology and particle physics, and has become an ancillary target of the CMB polarization experiments \cite{Xia:2007qs,Feng:2006dp,Ade:2015cao,Alighieri:2015doa,Gluscevic:2012me,Gruppuso:2015xza,Gruppuso:2011ci,Gubitosi:2012rg,Mei:2014iaa,Molinari:2016xsy,Aghanim:2016fhp,Contreras:2017sgi,Array:2017rlf}. The current upper bound on the constant rotation is $\sim 0.5$ deg \cite{Molinari:2016xsy,Contreras:2017sgi}. Constraints on the amplitude of the scale-invariant anisotropic rotation spectrum (defined in Eq.~(\ref{def:Aalpha})) are currently on the order of $0.1$ deg$^2$ \cite{Contreras:2017sgi,Array:2017rlf}. As we will show, future CMB experiments, such as LiteBIRD \cite{litebird}, Simons Observatory \cite{Ade:2018sbj}, a CMB-S4-like experiment \cite{Abazajian:2016yjj} and PICO \cite{Hanany:2019lle}, will improve these bounds by orders of magnitude.

The prospect of accurate measurements of CPR presents an opportunity for probing physics beyond the standard model, such as parity violating axion-photon interactions \cite{PecceiQuinn1977PRL,Weinberg1978PRL,Wilczek1978PRL}. These interactions result in different travel speeds of the two photon spin states, causing CPR \cite{Harari:1992ea,Carroll:1989vb,Carroll:1998zi,Pospelov:2008gg}. Axion-like parity-violating terms have been discussed in the context of inflation \cite{Freese:1990rb}, quintessence \cite{Wetterich:1987fm,Frieman:1995pm,Carroll:1998zi,Kaloper:2005aj,Dutta:2006cf,Liu:2006uh,Abrahamse:2007te,Lee:2013mqa}, baryogenesis \cite{Alexander:2016hxk} and neutrino number asymmetry \cite{Geng:2007va}. More generally, CPR probes violations of Lorentz invariance that could emerge in theories of quantum gravity \cite{Kostelecky:1988zi}, unconventional fields \cite{Kostelecky:2002ca,Bertolami:2003qs} and theories involving noncommutative spacetime \cite{Hayakawa:1999yt,Carroll:2001ws}. A general self-consistent description of Lorentz violation is provided by the Standard-Model Extension (SME) \cite{Colladay:1996iz,Colladay:1998fq,Kostelecky:2003fs}. Faraday Rotation (FR) by cosmic magnetic fields is another mechanism \cite{Kosowsky:1996yc}, where the rotation has a characteristic frequency dependence. As we will show, upcoming and future CMB experiments will significantly improve bounds on primordial magnetic fields (PMF) \cite{Yadav:2012uz,De:2013dra,Pogosian:2013dya,Pogosian:2018vfr}, providing an important observational handle on theories of inflation and the high energy universe \cite{Subramanian:2015lua}. 

The rest of the paper is organized as follows. We review the CPR formalism, the relevant observables and systematic effects in Section \ref{sec:rotation}. The forecasts for the future CMB polarization experiments, along with brief review of the current bounds, are presented in Section \ref{sec:forecast}. We elaborate on implications of improved bounds on CPR for fundamental physics in Section \ref{sec:physics}. We conclude with a summary in Section \ref{sec:summary}.

\section{Rotation of CMB polarization}
\label{sec:rotation}

Depending on the underlying physical mechanism (see Section \ref{sec:physics}), the CPR angle $\alpha$ could be a constant or a function of the line of sight, $\hat{n}$. In this Section, we briefly review the estimator used for both constant and anisotropic CPR.

CMB polarization maps are commonly separated into the so-called E- and B-modes~\cite{Kamionkowski:1996ks,Kamionkowski:1996zd,Seljak:1996gy}, which are the parity-even and parity-odd patterns of the polarization vector, which we will simply referr to as E and B. While E-modes are produced by Thomson scattering from intensity gradients at first order in cosmological perturbation theory, generating B requires sources with parity-odd components, such as gravitational waves \cite{Crittenden:1993ni}, topological defects \cite{Seljak:1997ii} or magnetic fields \cite{Seshadri:2000ky}. The weak lensing (WL) of CMB by large scale structures turns some of the E into B, generating the signal measured by POLARBEAR \cite{Ade:2014afa} and SPT \cite{Keisler:2015hfa}.

The CPR converts\footnote{The CPR discussed in this paper is restricted to rotation of linear polarization along the line of propagation. We do not consider circular polarization which is not expected to be present in the CMB \cite{Montero-Camacho:2018vgs}.} E into B, as well as B into E, although the latter effect is small enough to be ignored for very small rotation angles. Expanding $\alpha(\hat{n})$ into spherical harmonics, $\alpha(\hat{n}) = \sum_{LM} \alpha_{LM} Y_{LM}(\hat{n})$, the relation between the spherical expansion coefficients of the underlying E- and the induced B-mode can be written as \cite{Kamionkowski:2008fp,Gluscevic:2009mm}
\be
B_{lm}=2\sum_{LM}\sum_{l' m'}\alpha_{LM} E_{l' m'} 
\xi_{lml'm'}^{LM}H_{ll'}^L \ ,
\label{eq:blm}
\ee
where $\xi_{lml'm'}^{LM}$ and $H_{ll'}^L$ are related to Wigner $3$-$j$ symbols:
\ba
\nonumber
\xi_{lml'm'}^{LM} \equiv (-1)^m \sqrt{ (2l+1)(2L+1)(2l'+1) / 4\pi} \\
\times \left(
\begin{array}{ccc}
l  & L  & l'  \\
-m  & M  & m'    
\end{array}
\right) ; \ H_{ll'}^L \equiv 
\left(
\begin{array}{ccc}
l  & L  & l'  \\
2  & 0  & -2    
\end{array}
\right) \ ,
\ea
and the summation is restricted to {\it even} $L+l'+l$. In contrast, the WL conversion of E into B \cite{Hu:2001kj} couples the odd sums of the modes, making it orthogonal to the CPR effect. Eq.~(\ref{eq:blm}) also applies to the case of a constant CPR angle, in which case all $\alpha_{LM}$ are zero, except for $\alpha_{00}$.

\subsection{The mode-coupling estimator of the rotation}
\label{sec:estimator}

For an anisotropic $\alpha(\hat{n})$, Eq.~(\ref{eq:blm}) implies correlations between different multipoles of E and B. Since the CMB temperature (T) and E are correlated, CPR also correlates T with B.  The rotation angle can be extracted from EB and TB correlations \cite{Kamionkowski:2008fp,Gluscevic:2009mm,Yadav:2009eb}. Given measurements of B and E in frequency channels $i$ and $j$, respectively, the quantity 
\be
[{\hat \alpha}_{B^iE^j,LM}]_{ll'} = {2\pi \sum_{mm'} B^i_{lm}E_{l'm'}^{j*} \xi_{lml'm'}^{LM} \over (2l+1)(2l'+1)C_l^{EE}H_{ll'}^L}
\label{alphallpr}
\ee
provides an unbiased estimator of $\alpha_{LM}$ \cite{Pullen:2007tu,Yadav:2009eb,Gluscevic:2009mm,Gluscevic:2012me,Yadav:2012tn}. Note that $[{\hat \alpha}_{B^iE^j,LM}]_{ll'}$ is not symmetrical under interchange of $l$ and $l'$, and one should separately consider contributions from BE and EB correlations. Analogous quantities can also be constructed from products of T and B. Hence, given maps of T, E and B from a number of channels (labeled by indices $i,j$), one considers contributions from all quadratic combinations 
\be
A \in \{E^iB^j, B^iE^j, T^iB^j, B^iT^j\} \ .
\label{A-quad}
\ee 
The minimum variance estimator ${\hat \alpha}_{LM}$ is obtained by combining estimates from all $A$, accounting for the covariance between them. 

The variance in ${\hat \alpha}_{LM}$ was derived in \citet{Gluscevic:2009mm}. For a statistically isotropic CPR, it is defined as $\langle {\hat \alpha}^*_{LM} {\hat \alpha}_{L'M'} \rangle =\delta_{LL'} \delta_{MM'} [C_L^{\rm \alpha}+\sigma^2_{{\rm \alpha},L}]$, where $C_L^{\rm \alpha}$ is the CPR power spectrum that receives contributions from the sources of rotation (such as birefringence), while $\sigma^2_{{\rm \alpha},L}$ is the combined variance of individual estimators $[{\hat \alpha}_{B^iE^j,LM}]_{ll'}$. Using a notation similar to that in \cite{Gluscevic:2009mm}, we can write
\be
\sigma^{-2}_{{\rm \alpha},L} = \sum_{l' \ge l} G^L_{l l'} 
\sum_{A,A'} [({\cal C}^{l l'})^{-1}]_{AA'} \ 
Z_{l l'}^A Z_{l l'}^{A'} 
\ ,
\label{eq:ebnoise}
\ee
where the sum is restricted to even $l+l'+L$, $G^L_{l l'} \equiv (2l+1)(2l'+1) (H^L_{l l'})^2 / \pi$, $A$ and $A'$ label the relevant quadratic combinations of E, B and T listed in (\ref{A-quad}), 
\begin{align}
Z_{l l'}^{X^iB^j} = c^2 W^{ij}_{l l'} C^{XE}_l  , \\
Z_{l l'}^{B^iX^j} = c^2 W^{ij}_{l l'} C^{EX}_{l'} ,
\end{align}
with $X$ denoting either $T$ or $E$, and $W^{ij}_{l l'} \equiv \exp[-(l^2+l'^{2}) \theta^2_{ij}/16\ln 2]$ accounts for the finite width of the beam. We take $\theta_{ij} = \max[{\theta^i_{\rm fwhm},\theta^j_{\rm fwhm}}]$, where $\theta^i_{\rm fwhm}$ is the full-width-at-half maximum (FWHM) of the Gaussian beam of the $i$-th channel. 

The covariance matrix elements, $[{\cal C}^{l l'}]_{AA'}$, are
\begin{align}
\nonumber
[{\cal C}^{l l'}]_{X^iB^j,Y^kB^n}= {\tilde C}^{X^iY^k}_l {\tilde C}^{B^jB^n}_{l'} + \delta_{ll'} {\tilde C}^{X^iB^n}_l {\tilde C}^{B^jY^k}_{l'}\\
[{\cal C}^{l l'}]_{B^iX^j,B^kY^n}= {\tilde C}^{B^iB^k}_l {\tilde C}^{X^jY^n}_{l'} + \delta_{ll'} {\tilde C}^{B^iY^n}_l {\tilde C}^{X^jB^k}_{l'}
\label{eq:covar}
\end{align}
with $X$ and $Y$ standing for either E or T, and
\be
{\tilde C}^{X^iY^j}_l = C^{XY, {\rm prim}}_l+A_{\rm L}C^{XY, {\rm WL}}_l+ C^{XY, {\rm sys}}_l+ N^{X^iY^j}_l \ ,
\label{clvariance}
\ee
is the measured spectrum that includes the primordial contribution $C^{XY, {\rm prim}}_l$, the WL contribution $C^{XY, {\rm WL}}_l$, the systematic effects $C^{XY, {\rm sys}}_l$, and $N^{X^iY^j}_l$ that includes detector noise, assumed to be uncorrelated between the channels, and the residual contribution of galactic and atmospheric foregrounds. The de-lensing fraction $A_{\rm L}$ is introduced to account for the partial subtraction of the WL contribution. According to \cite{Hirata:2002jy}, the quadratic estimator method of \cite{Hu:2001kj} can reduce the WL contribution to ${\tilde C}_l^{BB}$ by a factor of $7$ (implying $A_{\rm L} =0.14$), with iterative methods promising a further reduction \cite{Hirata:2002jy}. 

The signal to noise ratio (SNR) of the detection of the CPR spectrum $C_L^{\rm \alpha}$ is given by 
\be
\left( S \over N \right)^2 = \sum_{L=1}^{L_{max}} {(f_{\rm sky}/2) (2L+1) [C_L^{\rm \alpha}]^2 \over [C_L^{\rm \alpha} + \sigma^2_{{\rm \alpha},L}]^2}.
\label{eq:ebsnrP}
\ee
The variance in the rotation estimator is the lowest at small $L$, since a rotation that is uniform over a large patch of the sky affects many E-modes in the same way. Hence, the main contribution to the SNR tends to come from the smaller $L$ modes, at least for the rotation spectra that are close to being scale-invariant.

As will be discussed in Sec.~\ref{sec:physics}, two special cases are of special interest: that of the scale-invariant rotation spectrum \cite{Pospelov:2008gg}, and that of the uniform rotation angle \cite{Carroll:1998zi}. The amplitude of the scale-invariant spectrum can be conveniently described by a constant parameter $A_\alpha$,
\be
A_\alpha \equiv {L(L+1) C_L^{\alpha} \over 2\pi}.
\label{def:Aalpha}
\ee
For a noise-dominated measurement, {\it i.~e} when $C_L^{\alpha} < \sigma^2_{\alpha,L}$ for all $L$, the SNR is proportional to $A_\alpha$. However, if $C_L^{\alpha}$ is larger than the variance for $L<L_S$, the contribution of these $L$ to the SNR is 
\be 
(S/N)^2 \approx \sum_{L=1}^{L_S} f_{\rm sky} (2L+1)/2 \approx L_S^2/2 \ ,
\ee
where $L_S$ is found by setting $C_{L_S}^\alpha = \sigma^2_{\alpha,L_S}$. For a scale-invariant spectrum, this implies $L_S(L_S+1) \approx L^2_S \approx 2 \pi A_\alpha/ \sigma^2_{\alpha,L}$, or $L_S \approx \sqrt{2\pi A_\alpha}/\sigma_{\alpha,L_S}$. This leads to
\be
{S \over N} \approx {\sqrt{\pi A_\alpha} \over \sigma_{\alpha,L_S}},
\label{eq:snsignal}
\ee
and implies a {\it linear} dependence of the SNR on the rotation angle, which is expected, since the rotation estimator can be used to directly reconstruct the rotation map. This translates into a linear dependence of the SNR on the parameters of the underlying theory for the cause of the rotation, such as the axion decay constant or the strength of the primordial magnetic field, which makes the CPR a sensitive probe of fundamental physics.

The case of the uniform rotation angle is a sub-case of a general rotation, with Eq.~(\ref{eq:ebnoise}) for the combined variance in the mode-coupling estimator remaining valid for $L=0$. For a full sky, the $00$th multipole of the rotation angle is related to the angle via $\alpha_{00}=\sqrt{4\pi} \alpha$. For a partial sky coverage, the variance in a uniform $\alpha$ is given by 
\be
\sigma^2_{\alpha} = {\sigma^2_{{\rm \alpha},0} \over 4\pi f_{\rm sky}} \ ,
\label{eq:sig_alpha}
\ee
where $\sigma^2_{{\rm \alpha},0}$ is given by Eq.~(\ref{eq:ebnoise}) with $L=0$. The uniform rotation generates parity-odd angular spectra of EB and TB type:
\be
C_l^{EB} = 2 \alpha C_l^{EE}  \ {\rm and} \ C_l^{TB} = 2 \alpha C_l^{TE} \ .
\ee
As explicitly shown in \cite{Gluscevic:2009mm}, the signal-to-noise of a detection of a uniform $\alpha$ (using the variance in (\ref{eq:sig_alpha})) is equivalent to the signal-to-noise of detection of the EB and TB spectra. We will use (\ref{eq:sig_alpha}) in our forecasts as it includes the covariance of multiple frequency channels.

\subsection{The B-mode spectrum induced by CPR}

In addition to mode-coupling correlations, CPR induces a contribution to the B-mode spectrum. Ignoring the effects of the finite width of the last scattering surface, the B-mode spectrum induced by CPR with a spectrum $C_L^{\alpha}$ is
\be
C_l^{BB(\alpha)} = \frac{1}{\pi} \sum_L  (2L+1) C^{\alpha}_L 
    \sum_{l_1}  (2l_1+1) C^{EE}_{l_1} (H_{ll_1}^L)^2 .
\label{eq:clbb_spec}
\ee
The signal-to-noise in detecting CPR from the measurement of the BB spectrum is given by
\be
\left( S \over N \right)^2 = \sum_{l=L_{min}}^{L_{max}} {f_{\rm sky} (2l+1) [C_l^{BB(\alpha)}]^2 \over 2[\tilde{C}_l^{BB} + C_l^{BB(\alpha)}]^2} \ ,
\label{eq:bbsnr}
\ee
where the covariance $\tilde{C}_l^{BB}$ includes instrumental noise, the systematic effects and the WL contribution. 

For a constant CPR angle $\alpha$, we have
\be
C_l^{BB(\alpha)} = \alpha^2 C^{EE}_l .
\label{eq:clbb_const}
\ee
and the signal-to-noise becomes
\be
\left( S \over N \right)^2 = \sum_{l=L_{min}}^{L_{max}} {f_{\rm sky} (2l+1) [\alpha^2 C_l^{EE}]^2 \over 2 [\tilde{C}_l^{BB} + \alpha^2 C_l^{EE}]^2} .
\label{eq:bbsnr_const}
\ee
Note that the signal-to-noise in BB is quadratic in $\alpha$, while in the case of EB and TB correlations it is essentially linear in CPR. Thus, given a CMB experiment of sufficiently low noise and high resolution, the latter offer a more sensitive probe of the CPR than the B-mode spectrum. This is true for both constant and anisotropic rotation. 

\subsection{Beam systematics}
\label{sec:beam}

Optical imperfections in the telescope itself, known as ``beam systematics'', are capable of generating spurious CPR \cite{Shimon:2007au,Pagano:2009kj,Gubitosi:2012rg,Su:2009tp,Su:2010wa} due to the leakage of power from the ``standard'' correlations of TT, TE and EE type.  Beam systematics generate non-zero parity-odd angular spectra $C^{EB}_l$ and $C^{TB}_l$ in addition to contributing to all parity-even spectra, including $C^{BB}_l$.  Their multipole-dependence can be modelled \cite{Shimon:2007au,Su:2010wa}, allowing one to partially separate this non-cosmological signal from the data. The separation cannot be perfect because contributions from the systematics increase the variance and because of the uncertainties associated with the beam model. Moreover, the pixel-rotation systematic effect caused by the misalignment of the telescope is fully degenerate with a uniform (across the sky) CPR angle\footnote{In fact, measured TB and EB correlations are sometimes used to correct for the misalignment of the telescope based on the assumption that the underlying CPR is vanishing \cite{Yadav:2012tn,Keating:2012ge, Kaufman:2013vbd}.}.

In our forecasts, we use the parameterized forms of $C^{XY, {\rm sys}}_l$ derived in \cite{Shimon:2007au} for the spurious contributions to CMB spectra caused by the differential pointing, differential ellipticity, and differential rotation in a dual polarized beam. In the formalism of \cite{Shimon:2007au}, these three systematic effects are controlled by the corresponding parameters $\rho$, $e$, and $\epsilon$, which, for simplicity, were assumed to be independent of $l$. Under this simplifying assumption, beam systematics contribute to Eq.~(\ref{eq:covar}) only at $l=l'$. For each experiment, we evaluate $C^{XY, {\rm sys}}_l$ assuming the uncertainties in $\rho$, $e$ and $\epsilon$ are reduced to sufficiently low levels that allow the experiment to achieve its scientific targets, {\it i.e.} to exhaust its nominal capacity to detect B-mode polarization. Specifically, two requirements must be satisfied:
\begin{enumerate}
\item Beam systematics should not reduce the experiment's ability to measure $r$; and
\item Beam systematics should allow the experiment to measure the lensing B-mode on relevant scales, {\it i.e.} we require that $C^{BB, {\rm sys}}_{l*} < N^{BB}_{l*}$ for $l<l_*$, where $l_*=1150$ corresponds to the peak of the WL spectrum.
\end{enumerate}

When modelling the differential pointing effect, we assume that one of the beams has no pointing error, while the other beam has a pointing error $\rho$. The angle $\theta$ of the second beam is a free parameter that we fix at $\theta=45^\circ$. Our calculation of the quadrupole effect assumes that the two beams have the same ellipticity $e$, and the angles that the polarization axes make with the major axes of the two beams are taken to be $\psi_1=45^\circ$ and $\psi_2=0$. The values of $\rho$, $e$ and $\epsilon$, derived for each experiment under the above assumptions are given in Table~\ref{tab:exp}, and we refer the reader to \cite{Shimon:2007au} for the complete description of the beam model.

\section{Bounds on the rotation of CMB polarization}
\label{sec:forecast}

\begin{table*}[tbp]
\begin{tabular}{|c||c|c||c|c||c|c||c|c|}
\multicolumn{1}{c||}{} 
          &  \multicolumn{2}{|c||}{LiteBIRD}           & \multicolumn{2}{|c||}{SO SAT}    &  \multicolumn{2}{|c||}{CMB-S4-like}    & \multicolumn{2}{|c}{PICO} \\
\multicolumn{1}{c||}{} 
          & \multicolumn{2}{|c||}{space}     & \multicolumn{2}{|c||}{ground}  &  \multicolumn{2}{|c||}{ground}   & \multicolumn{2}{|c}{space} \\
\multicolumn{1}{c||}{} 
          & \multicolumn{2}{|c||}{$f_{sky}=0.6$}     & \multicolumn{2}{|c||}{$f_{sky}=0.1$}  & \multicolumn{2}{|c||}{$f_{sky}=0.4$}   & \multicolumn{2}{|c}{$f_{sky}=0.6$} \\    
\hline            
\multicolumn{1}{|c||}{ target sensitivity\footnote{The expected 68\% CL upper bound on $r$ assuming it is undetectably small.} to $r$ } 
        & \multicolumn{2}{|c||}{$\sigma_r=0.001$}     & \multicolumn{2}{|c||}{$\sigma_r=0.003$}  & \multicolumn{2}{|c||}{$\sigma_r=0.0005$}   & \multicolumn{2}{|c|}{$\sigma_r=5\times 10^{-5}$} \\ 
\hline
\multicolumn{1}{|c||}{de-lensed fraction\footnote{Perfect de-lensing corresponds to $f_L=0$, no de-lensing to $f_L=1$.} }
        & \multicolumn{2}{|c||}{$f_L=0.5$}     & \multicolumn{2}{|c||}{$f_L=0.5$}  & \multicolumn{2}{|c||}{$f_L=0.15$}   & \multicolumn{2}{|c|}{$f_L=0.1$} \\              
\hline
\multicolumn{1}{|c||}{($10^4 \rho$, $10^3 e$, $10^2 \epsilon$)\footnote{The beam systematics parameters used in our forecasts as described in Sec.~\ref{sec:beam}.}}
        & \multicolumn{2}{|c||}{(4, 1.5, 1.5)}     & \multicolumn{2}{|c||}{(1, 0.2, 1.5)}  & \multicolumn{2}{|c||}{(2, 50, 0.4)}   & \multicolumn{2}{|c|}{(1, 1.5, 0.4)} \\              
\hline
Frequency   & $\theta_{\rm fwhm}$ &  $\sigma_P$ & $\theta_{\rm fwhm}$ &  $\sigma_P$ & $\theta_{\rm fwhm}$ &  $\sigma_P$   & $\theta_{\rm fwhm}$  &  $\sigma_P$ \\
(GHz) &  $'$ 	  & $\mu$K-$'$&  $'$     &  $\mu$K-$'$ &  $'$	   &  $\mu$K-$'$ &  $'$  &  $\mu$K-$'$ \\
\hline
 60 	 & 48       & 19.5         				 & -        & -        			& -         & -         	& $13$  & $3.9$ \\
 70    & 43        & 15.8       					& -        & -       	        	                 & -         & -         	& -  & - \\
 75    & -        & -         					& -        & -       	  			& -         & -         	& $11$  & $3.2$ \\
 78    & 39        & 13.3         				& -        & -       	     			& -         & -         	& -  & - \\
 90    & 35       & 11.5        					& -        & -        				& -         & -         	& $9.5$  & $2$ \\
95     & -         &    -    					& $30$ & $2.7$ 		& $2.2$ & $2.1$		& -         &    -  \\
100   & 29      &  9.0    					& -        & -				& -         & -			& -         &    -  \\
110   & - 	      	&   -    					& -        & -				& -         & -			& $7.9$    & $1.7$ \\
120   & 25  &  7.5   						& -        & -				& -         & -			-    & - \\
130   & -	    & -	    					& -        & -				& -         & - 			& $7.4$    & $1.6$ \\
140   & 23     & 5.8    					& -        & -				& -         & -			& -         & -   \\
150   & -	 	& -						& $17$ & $3$ 	       		& $1.5$    & $2.1$ 		& $6.2$    & $1.4$ \\
\hline
\end{tabular}
\caption{\label{tab:exp} Parameters of the CMB experiments considered in our forecast. The values of frequencies in the first column are rounded. Channels with frequencies below 60 and above 150 GHz are dominated by the galactic foregrounds and are not included.}
\end{table*}
 
Using the formalism presented in the previous section, we perform a forecast of expected bounds on the CPR for the following upcoming and proposed experiments:
\begin{itemize}
\item \emph{Lite (Light) satellite for the studies of B-mode polarization and Inflation from cosmic background Radiation Detection (LiteBIRD)} \cite{litebird} -- a proposed small satellite observatory, with channels covering a wide range of frequencies, targeting B-modes in the $2 < \ell < 200$ range, and aiming to constrain the tensor-to-scalar ratio $r$ at a level of $\sigma_r=0.001$.  In our forecasts, we only include channels in the 60-150 GHz range, where the galactic foregrounds are relatively weak. For LiteBIRD, we assume that the residual foreground contribution is equal to the noise in the lowest noise channel;
\item \emph{Simons Observatory (SO)} \cite{Ade:2018sbj}, a ground-based experiment currently under development, consisting of one 6 meter Large Aperture Telescope (LAT) and three 0.5 meter Small Aperture telescopes (SAT), aiming to achieve $\sigma_r=0.003$. For SO, we assume the SAT parameters and the forecasted noise curves from \cite{Ade:2018sbj}\footnote{The SO noise curves are available at \url{https://simonsobservatory.org/assets/supplements/20180822_SO_Noise_Public.tgz}} that include modelling of atmospheric and galactic foregrounds;
\item \emph{a Stage IV ground based experiment like CMB-S4} \cite{Abazajian:2016yjj} covering 40\% of the sky at 95 and 150 GHz with $\sim 1$ arcmin resolution and noise levels of $\sim 1$ $\mu$K-arcmin. For CMB-S4-like, we assume that the residual foreground contribution is equal to the noise in the lowest noise channel;
\item \emph{Probe of Inflation and Cosmic Origins (PICO)} \cite{Hanany:2019lle}, a proposed mid-cost space mission mapping the full sky using multiple channels covering a wide range of frequencies at a resolution of a several arcmin and noise levels of $\sim 1$ $\mu$K-arcmin. In our forecasts, we consider channels in the 60-150 GHz range and use the forecasted noise curves from \cite{hill:private} that were used in \cite{Hanany:2019lle} and include modelling of atmospheric and galactic foregrounds using the methodology described in \cite{Ade:2018sbj}.
\end{itemize}
The parameters assumed for each of the experiments are summarized in Table~\ref{tab:exp}.

We consider rotation by both a uniform and an anisotropic rotation angle, quantifying the results in terms of the expected 68\% confidence level (CL) bounds on the following parameters:
\begin{itemize}
\item the constant rotation angle $\alpha$, in arcmin ($'$);
\item the amplitude of the scale invariant rotation spectrum $A_\alpha$ of Eq.~(\ref{def:Aalpha}), in deg$^2$;
\item the quadrupole moment of the rotation, $\sqrt{C_2^\alpha/4\pi}$, in arcmin, assuming a scale-invariant spectrum:
\be
\sqrt{C_2^\alpha \over 4\pi} = \sqrt{A_\alpha \over 12}.
\label{eq:C2}
\ee
\end{itemize}
In addition, we plot the statistical uncertainty in $C_L^\alpha$, under the assumption of no CPR,
\be
\sigma_{C_L^\alpha} = {\sigma^2_{\alpha,L} \over \sqrt{f_{\rm sky}(2L+1)/2}}
\label{eq:varianceCL}
\ee
as a function of $L$, in deg$^2$. Before presenting the forecasts, we briefly review the current bounds on CPR.

\subsection{Current bounds on CPR}

The current bound on the uniform rotation angle is $\alpha < 0.5^\circ$ at 68\% CL derived by Planck \cite{Aghanim:2016fhp} from the upper limit on parity-odd two-point correlations of EB and TB type (see also \cite{Xia:2009ah,Gruppuso:2015xza,Molinari:2016xsy,Contreras:2017sgi}). It improved on the 68\% CL bound of $\alpha < 1.5^\circ$ from WMAP7 \cite{Komatsu:2010fb}.

The existing constraints on the anisotropic rotation are based on the assumption of a scale-invariant rotation spectrum. The present bound is $A_\alpha < 0.07$ deg$^2$ at 95\% CL obtained in \cite{Contreras:2017sgi} using a pixel based approach to directly estimate the rotation angle on local patches of the Planck polarization maps. According to the scaling in Eq.~(\ref{eq:snsignal}), the corresponding 68\% CL bound would be approximately $0.02$ deg$^2$.  Expressed in terms of the quadrupole anisotropy, the bound is $\sqrt{C_2/4\pi} < 5'$ at 95\% CL or approximately $3'$ at 68\% CL.  A comparable bound, $A_\alpha < 0.11$ deg$^2$ at 95\% CL, was obtained by BICEP2/Keck \cite{Array:2017rlf} using the mode coupling estimator introduced in the previous section. Prior bounds on $A_\alpha$ were also derived from WMAP7 \cite{Gluscevic:2012me} ($A_\alpha < 12$ deg$^2$ at 68\% CL) and POLARBEAR \cite{Ade:2015cao} ($A_\alpha < 1$ deg$^2$ at 95\% CL). Bounds on $A_\alpha$ from POLARBEAR and SPTPol B-mode spectra were also derived in \cite{Liu:2016dcg}. The present day bounds are summarized in Table~\ref{tab:rotation}.

\begin{table*}[tbp]
\centering
\begin{tabular}{c|c||c|c|c||c|c|c||c|c|c||c|c|c||c|c|c}
  \multicolumn{2}{c||}{}  & \multicolumn{3}{c||}{Current} &  \multicolumn{3}{c||}{LiteBIRD} & \multicolumn{3}{c||}{SO} & \multicolumn{3}{c||}{CMB-S4-like} & \multicolumn{3}{c}{PICO} \\
\hline
DL & BS & $\alpha$&  $A_\alpha$ &  $\sqrt{C_2^\alpha \over 4\pi}$
& $\alpha$ &  $A_\alpha$  &  $\sqrt{C_2^\alpha \over 4\pi}$
 & $\alpha$  &  $A_\alpha$  & $\sqrt{C_2^\alpha \over 4\pi}$ & $\alpha$  &  $A_\alpha$  &  $\sqrt{C_2^\alpha \over 4\pi}$ & $\alpha$  &  $A_\alpha$ &  $\sqrt{C_2^\alpha \over 4\pi}$ \\
 &  & $'$ & $10^{-2}$deg$^2$ & $'$
& $'$ & $10^{-3}$deg$^2$ &  $'$
 & $'$ &   $10^{-4}$deg$^2$  & $'$  & $'$ &  $10^{-5}$deg$^2$ & $'$  &  $'$ &  $10^{-5}$deg$^2$ & $'$  \\
\hline
yes & no   & -        & -          & -  &   1.3 &  2.7 & 0.9  &   0.56 & 3 &  0.29 &    0.1 & 1.4 & 0.065 &      0.05  & 0.4 & 0.035  \\
yes & yes & -         & -          & -  &  1.5 &  3.3 & 1.0  &   0.66 & 4   &  0.35 &      0.11 & 2.0 & 0.08  &     0.06  & 0.5 & 0.04   \\
no   & no   & -        & -          & -  &  1.4 & 	3.5 & 1.0  &   0.64 & 5.0 &  0.4 &    0.13 & 2.5 & 0.09  &     0.08  & 1.2  & 0.06 \\
no   & yes  & 30     & 2         & 3   & 1.6  & 	4.0 & 1.1  &   0.71 & 5.5 &  0.4  &  0.15 & 3.3 & 0.1   &     0.09  & 1.4 & 0.065 \\
\hline
\end{tabular}
\caption{\label{tab:rotation} Current and forecasted $68$\% CL bounds on the uniform and the anisotropic CPR parameters.
} 
\end{table*}

\subsection{Forecasted CPR bounds}

\begin{figure*}[tbp]
\includegraphics[width=1.3\columnwidth,angle=270]{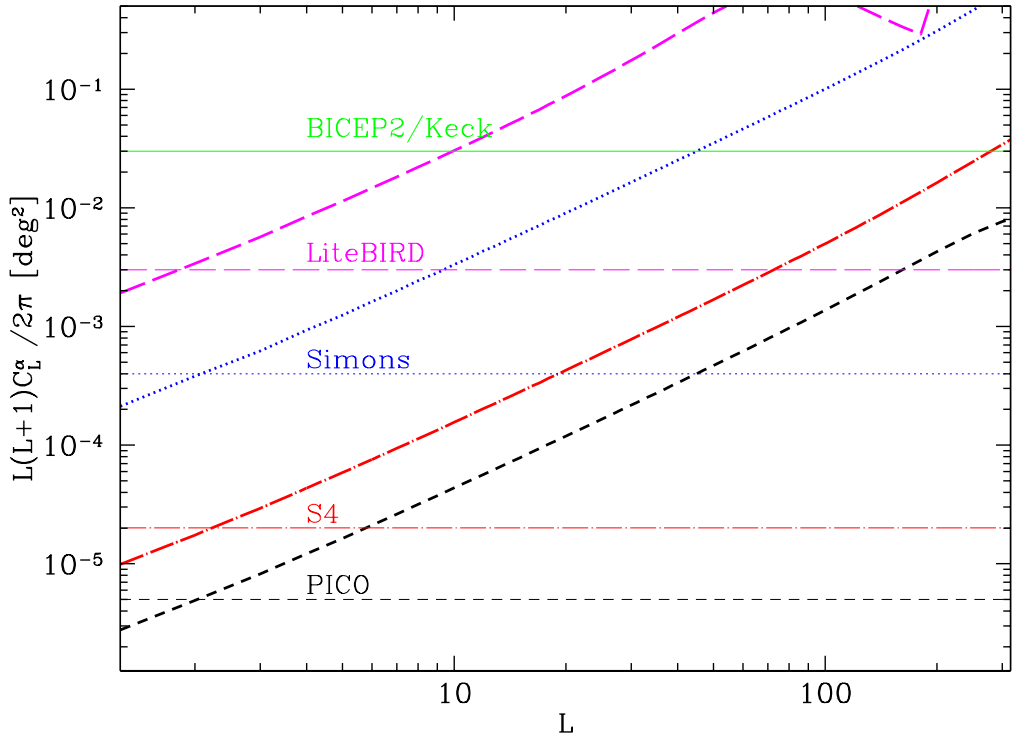}
\caption{The thick lines show the statistical uncertainty in $C_L^{\alpha}$, given by Eq.~(\ref{eq:varianceCL}), forecasted for the four experiments considered in this work. These curves assume de-lensing by a fraction $f_L$ given for each experiment in Table~\ref{tab:rotation}, and account for the effects of beam systematics. The thinner horizontal lines indicate the corresponding expected 68\% CL bounds on the amplitude of the scale-invariant rotation spectrum $A_\alpha$. The thin green solid line shows the current bound on $A_\alpha$ from BICEP2/Keck \cite{Array:2017rlf}.}
\label{fig:varianceCL}
\end{figure*}

Table~\ref{tab:rotation} summarizes the forecasted bounds on the uniform and the anisotropic CPR expected from the four experiments considered in this work. 

The ability of a given experiment to constrain CPR is determined primarily by its resolution and the effective noise that includes the residual foreground contributions. Specifically, an optimal experiment for detecting a scale-invariant rotation spectrum would have the resolution to measure most of the $\ell$-modes around the peak of the E-mode spectrum, or $500 \lesssim \ell \lesssim 3000$. Having better resolution does not significantly improve constraints on the rotation simply because there is less power in the polarization on smaller scales. However, if the rotation spectra were not scale-invariant but had a significant blue tilt, with most of the power on small scales, having polarization measurements at a higher resolution could be beneficial. We leave investigation of this latter possibility for future work.

From Table~\ref{tab:rotation} one can see that LiteBIRD would lower the bounds on CPR by an order of magnitude, while the Simons Observatory will lower them by two orders. Both CMB-S4-like and PICO are capable of improving them by yet another order of magnitude, with PICO being somewhat more sensitive to CPR thanks to the lower detector noise.

In Fig.~\ref{fig:varianceCL} we plot the forecasted statistical uncertainty in the rotation spectrum $C_L^{\alpha}$ given by Eq.~(\ref{eq:varianceCL}). The curves take into account the contribution of beam systematics, and assume partial de-lensing by a fraction $f_L$, given for each experiment in Table~\ref{tab:exp}. The plot also shows the forecasted 68\% CL bounds on the amplitude of the scale-invariant rotation spectrum $A_\alpha$ (the horizontal lines) for each experiment, along with the current bound on $A_\alpha$ from BICEP2/Keck \cite{Array:2017rlf}.

\section{Implications for fundamental physics}
\label{sec:physics}

\subsection{A pseudoscalar field coupled to electromagnetism}
\label{sec:axion}

A number of well-motivated extensions of the standard model involve a (nearly) massless axion-like pseudoscalar field coupled to photons via the Chern-Simons (CS) interaction term. The relevant contribution to the Lagrangian can be written as
\be 
\mathcal{L}_{a\gamma}=-{1\over 4}F_{\mu\nu}F^{\mu\nu}+{1\over 2}\partial_\mu a\partial^\mu a + \frac{a}{2f_a}F_{\mu\nu}\tilde{F}^{\mu\nu} -\left( {1\over 2}{m_a^2} a^2\right),
\label{eq:pseudophi}
\ee
where $F_{\mu\nu}$ is the electromagnetic field strength, $\tilde{F}_{\mu\nu}$ is its dual, $a({\bf x},\eta)$ is the pseudoscalar field, $f_a$ is the axion decay rate, and $m_a$ is the axion mass which is either zero or constrained to be very small. One should think of $a$ as being the phase of a complex scalar field with a spontaneously broken $U(1)$ symmetry, {\it i.e.} a (pseudo-) Goldstone boson, with the value of $f_a$ set by the symmetry breaking scale.  Axions were first introduced in the context of the QCD \cite{PecceiQuinn1977PRL,Weinberg1978PRL,Wilczek1978PRL} as a solution of the strong CP violation problem \cite{tHooft:1976rip}. Axion-like fields are ubiquitous in string theory \cite{Svrcek:2006yi,Fox:2004kb} and can be relevant in developing models of inflation \cite{Freese:1990rb}, quintessence \cite{Wetterich:1987fm,Frieman:1995pm,Carroll:1998zi,Kaloper:2005aj,Dutta:2006cf,Abrahamse:2007te}, baryogenesis \cite{Alexander:2016hxk} and neutrino number asymmetry \cite{Geng:2007va}. We refer the reader to \cite{Marsh:2015xka} for a recent review of cosmological implications of axion-like fields.

The parity-violating term in (\ref{eq:pseudophi}) makes the right- and left-handed polarization states propagate at different velocities, 
\be
\ddot{A}_{\pm}(k,\tau)+\left[ k^{2} \pm {2k \over f_a} (\dot{a} + {\hat n} \cdot {\vec \nabla} a) \right] A_{\pm}(k,\tau)=0,
\ee
where the vector-potential is decomposed into $A_{\pm}=A_{x}\pm iA_{y}$, a phenomenon known as birefringence. This causes a rotation of the linear polarization of an electromagnetic wave as it propagates \cite{Harari:1992ea}. If the wavelength of the radiation is much smaller than the typical scale over which $a$ varies, the rotation angle is independent of the wave's frequency and is given by $\Delta \alpha = \Delta a /f_a$, where $\Delta a$ is the net change in $a$ along the photon's trajectory \cite{Harari:1992ea,Carroll:1989vb,Carroll:1998zi,Pospelov:2008gg}. In order to produce any rotation of the CMB polarization, the axion mass must be smaller than the Hubble scale at decoupling,
\be
m_a < H_{\rm dec} \sim 10^{-28} \ eV,
\label{eq:ma_up}
\ee
otherwise, the axion will start oscillating around the minimum of the potential giving $\Delta a=0$. Note that the same criterion prevents $a$ from being the dark matter, since being a matter particle of relevance to structure formation requires it to start oscillating prior to decoupling.

A uniform CPR angle is possible if a) $\dot{a}$ is non-zero between the time of decoupling and today, and b) the average value of $a$ is non-zero at decoupling. The first of these conditions requires the mass to be sufficiently large for $a$ to be dynamical between the decoupling and today \cite{Carroll:1998zi}, namely, $m_a > H_0 \sim 10^{-33}$ eV. The second condition requires the value of $a$ to be uniform across the universe, which would be the case if its value was set during or prior to inflation. More specifically, the initial value of $a$ is set randomly at the time of the $U(1)$ symmetry breaking. If inflation happened at a scale $H_I>f_a$, the sky-averaged value of $a$ would be zero, as it would correspond to averaging over its value in many causally disconnected parts of the universe. On the other hand, if $f_a \ge H_I$, our observable universe would originate from the same patch that was causally connected at the time of symmetry breaking and the initial value of $a$ would be uniform across the sky. Hence, having an observable uniform CPR angle requires $f_a>H_I$ in addition to $10^{-33} {\rm eV} \lesssim m_a \lesssim 10^{-28}$ eV. 

As discussed in Sec.~\ref{sec:estimator}, a uniform CPR angle would manifest itself in non-zero $C_\ell^{TB}$ and $C_\ell^{EB}$ \cite{Lue:1998mq} and imply a global violation of parity in the universe. As our forecasts have shown, future experiments will improve the sensitivity to a constant CPR by over two orders of magnitude. In the context of axion-like fields, they will bound $\dot{a}/f_a$ and translate into constraints on a combination of $m_a$ and $f_a$ that would be complimentary to those from axion dark matter searches. In particular, a detection of a uniform CPR angle would imply a non-zero axion mass. We further discuss the uniform rotation angle in Sec.~\ref{sec:lvp} in the context of a general framework of searching for Lorentz violating extensions of the standard model.

Generally, the pseudoscalar $a$ would vary in space and time, with the spatial distribution largely determined by whether the symmetry breaking scale is above or below that of inflation. If $f_a<H_I$, then $a({\bf x},\eta)$ is expected to be uncorrelated on scales larger than the horizon size at the time of the symmetry breaking, implying a blue rotation spectrum on scales probed by CMB experiments with a cut off at an extremely high value of $L$. Such a CPR spectrum would have practically no power at low $L$ and would be undetectable. However, the breaking of the  $U(1)$ symmetry would also produce global cosmic strings \cite{Vilenkin:1982ks} which would remain topologically stable up to the epoch corresponding to the very small axion mass scale $m_a$ (which could be zero). A scaling network of axion strings would act as a continuous source of perturbations, sourcing axion fluctuations on scales corresponding to the horizon at any given time \cite{Yamaguchi:1998iv,Klaer:2017qhr,Gorghetto:2018myk,Kawasaki:2018bzv}. Detailed properties of such a spectrum and its effect on the CPR could be a subject of a future investigation.

Of special interest is the case when $f_a>H_I$, in which case stochastic fluctuations in the pseudoscalar field would be generated during the period of inflation \cite{Lyth:1991ub,Lyth:1992tx,Fox:2004kb}. This would result in a scale-invariant spectrum of the CPR angle with an amplitude
\be
A_\alpha = \left({H_I \over 2\pi f_a}\right)^2.
\ee
Thus, an upper bound on $A_\alpha$ implies a lower bound on the coupling scale $f_a$. In \cite{Pospelov:2008gg}, the authors studied the CMB B-mode spectrum generated by such a CPR (see Eq.~(\ref{eq:clbb_spec})) and derived a 95\% CL upper bound of $A_\alpha < 4.2 \times 10^{-3} \ {\rm rad}^2 = 13.8 \ {\rm deg}^2$ from the upper bound on the BB spectrum from QUaD \cite{Pryke:2008xp}, implying $f_a > 2.4 \times 10^{14} \ {\rm GeV} \times H_{14}$, where $H_{14}=H_I/10^{14}$GeV. As discussed in Section~\ref{sec:rotation}, given a CMB experiment of sufficiently low noise and high resolution, the mode-coupling EB and TB correlations offer a more sensitive probe of the rotation angle compared to the BB spectrum. The present 95\% CL bound on $A_\alpha$ from BICEP2/Keck \cite{Array:2017rlf} is $0.11$ deg$^2$, corresponding to $f_a > 2.7 (5.3) \times 10^{15} \ {\rm GeV} \times H_{14}$ at 95\% (68\%) CL.

The current and future 68\% CL lower CMB bounds on $f_a$ are shown in Table~\ref{tab:fa}. They are significantly tighter than those obtained from astrophysical probes of pseudoscalar interactions \cite{Galaverni:2014gca,Kamionkowski:2010ss,Loredo:1997gn}, assuming that the inflationary scale is not significantly below $14$ GeV. Generally, low-mass particles such as neutrinos and axions would be produced in the interior of stars, and stellar constraints typically require $f_a > 10^{11}$ GeV \cite{Raffelt:1999tx}.  The bound obtained by the CERN Axion Solar Telescope (CAST) experiment, which searched for the direct emission of pseudoscalars from the solar interior, is $f_a > 2 \times 10^{10}$ GeV \cite{Andriamonje:2007ew}. The bounds from laboratory experiments, such as the Polarization of Vacuum with LASer (PVLAS) experiment \cite{DellaValle:2015xxa}, are significantly weaker than those from astrophysics. 

Experiments such as CMB-S4-like and PICO are able to probe $f_a \sim {\rm a \ few} \times 10^{17} \ {\rm GeV} \times H_{14}$, in the range close to the Planck scale of $10^{19}$ GeV. In particular, this would exclude the range of $f_a \sim 10^{16} \ {\rm GeV} \times H_{14} $ that is of most interest for string theory, leading to non-trivial bounds on the string theory axions \cite{Svrcek:2006yi,Pospelov:2008gg} and implementations of inflation in the related models.

\begin{table}[tbp]
\centering
\begin{tabular}{c|c|c|c|c|c}
& Current &  LiteBIRD & SO & CMB-S4-like & PICO \\
\hline
$f_a$ [$H_I$] & 50 & 200  &  500   & 2000 & 4000  \\
 \hline
\end{tabular}
\caption{\label{tab:fa} Current and forecasted $68$\% CL lower bounds on the axion decay constant $f_a$, in the units of the energy scale of inflation, $H_I = 10^{14}$ GeV $\times H_{14}$. These are inferred from the bounds on $A_\alpha$ that include the effects of beam systematics and de-lensing.
} 
\end{table}

\subsection{Faraday Rotation by a Primordial Magnetic Field}
\label{sec:pmf}

The origin of micro-Gauss ($\mu$G) strength galactic magnetic fields is one of the long standing puzzles in astrophysics \cite{Widrow:2002ud}. Producing them with a dynamo mechanism requires a seed field of a certain minimum strength \cite{Widrow:2011hs}. Adding to the puzzle is the presence of $\mu$G strength fields in proto-galaxies too young to have gone through the number of revolutions necessary for the dynamo to work \cite{Athreya:1998}. There is also preliminary evidence for lower limits on primordial magnetic fields (PMF) from observations of cosmic rays for magnetic fields in the intergalactic space coherent over cosmological distances \cite{Neronov:1900zz,Takahashi:2011ac,Dolag:2010ni,Tavecchio:2010ja,Taylor:2011bn,Vovk:2011aa}. PMFs could have been generated in the aftermath of phase transitions in the early universe \cite{Vachaspati:1991nm}, during inflation \cite{Turner:1987bw,Ratra:1991bn}, or at the end of inflation \cite{DiazGil:2007dy}. Once produced, they would be sustained by the primordial plasma in a frozen-in configuration until the epoch of recombination and beyond leaving potentially observable imprints in the CMB. Thus, improved constraints on the PMF are valuable tools for discriminating among different theories of the early universe \cite{Barnaby:2012tk,Long:2013tha,Durrer:2013pga}.

A stochastic PMF contributes to the CMB anisotropy through metric perturbations and the Lorentz force exerted on ions in the pre-recombination plasma \cite{Seshadri:2000ky,Mack:2001gc,Lewis:2004ef,Finelli:2008xh,Paoletti:2008ck,Shaw:2009nf}. It also generates Faraday rotation (FR) of CMB polarization converting E modes into B modes \cite{Harari:1996ac,Scoccola:2004ke,Kahniashvili:2008hx,Pogosian:2011qv} and inducing mode-coupling correlations between E, B and T \cite{Yadav:2012uz,Pogosian:2013dya}. CMB signatures depend on the shape of the PMF spectrum, which in turns is determined by the generation mechanism of the PMF. The originally proposed simple inflationary models of magnetogenesis  \cite{Turner:1987bw,Ratra:1991bn} predict a scale-invariant spectrum, although other values are possible in more complicated models \cite{Bonvin:2013tba}. Magnetic fields produced in phase transitions after inflation have blue spectra with most power on very small scale. We will focus on the well-motivated case of the scale-invariant PMF \cite{Kahniashvili:2016bkp} that is most likely to have observable CPR \cite{Pogosian:2011qv,Yadav:2012uz}.

It is conventional to quote limits on the PMF in terms of $B_{\lambda}$, which is the magnetic field strength smoothed over a region of comoving size $\lambda$. For a scale-invariant PMF, this measure is independent of $\lambda$, and is the same as the effective PMF strength obtained by taking the square root of the magnetic energy density \cite{Mack:2001gc}, so we will quote the bounds in terms of $B_{\rm SI}=B_{\rm eff}=B_{\lambda}$. The current bound, derived from a combination of the 2015 Planck TT, EE, TE spectra \cite{Adam:2015rua} and the SPT B-mode spectrum \cite{Keisler:2015hfa} is $B_{\rm SI}<1.2$ nG at 95\% CL, or $<1$ nG at 68\% CL \cite{Zucca:2016iur}. In particular, the measured B-modes by SPT at small scales play an important role, reducing the bound on $B_{\rm SI}$ by a factor of two.

Because the magnetic contribution to CMB spectra scales as $B^4_{\rm SI}$, an orders-of-magnitude improvement in the accuracy of the B-mode spectrum would only result in a modest reduction of the bound on $B_{\rm SI}$. In contrast, Faraday Rotation (FR) scales linearly with $B_{\rm SI}$, promising much tighter bounds on the PMF \cite{Pogosian:2018vfr}. At present, such FR-based PMF bounds are not competitive compared to those from CMB spectra, {\it e.g.} the POLARBEAR collaboration obtained $B_{\rm SI} < 93$ nG at 95\% CL \cite{Ade:2015cao} based on the analysis of mode-coupling EB correlations in their 150\,GHz map, but they will improve dramatically with the lower noise and higher resolutions of future experiments.

To extract the FR signal, one can use the rotation angle estimator (\ref{alphallpr}) after accounting for the $\nu^{-2}$ frequency dependence of the FR angle $\alpha(\hat{n})$. Namely, one can use a combination of channels to constrain the frequency independent rotation measure (RM), defined as
\be
{\rm RM}(\hat{n}) \equiv c^{-2} \nu^2 \alpha(\hat{n}).
\ee
The details of constructing the multi-frequency RM estimator can be found in \cite{Pogosian:2013dya}.

A scale-invariant PMF implies a scale-invariant RM spectrum \cite{Pogosian:2011qv}, {\it i.~e.} the quantity 
\be
A^2_{\rm RM} = L(L+1)C_L^{\rm RM} / 2\pi
\label{a2rm}
\ee 
is constant over the scales of interest and is related to $B_{\rm SI}$ via \cite{De:2013dra} 
\be
A_{\rm RM}  \approx 50 \ {\rm rad/m^2} \ B_{\rm SI}/{\rm nG} \ .
\ee
The SNR of the detection of the primordial RM spectrum $C_L^{\rm RM,PMF}$ is given by 
\be
\left( S \over N \right)^2 = \sum_{L=1}^{L_{max}} {(f_{\rm sky}/2) (2L+1) [C_L^{\rm RM,PMF}]^2 \over [C_L^{\rm RM,PMF} + f_{\rm G}C_L^{\rm RM,G}+\sigma^2_{{\rm RM},L}]^2} \ ,
\label{eq:PMFsnr}
\ee
where $\sigma^2_{{\rm RM},L}$ is the variance in the RM estimator analogous to $\sigma^2_{{\alpha},L}$ that takes into account de-lensing and beam systematics, and $f_{\rm G}$ is the fraction of the Milky Way RM spectrum $C_L^{\rm RM,G}$ that may be known from other sources and can be subtracted. We use estimates of $C_L^{\rm RM,G}$ from \cite{De:2013dra} based on the galactic RM map of \cite{Oppermann:2011td}.

\begin{table}[tbp]
\centering
\begin{tabular}{c||c|c|c|c|c}
$B_{\rm SI}$ [nG] & Current &  LiteBIRD & SO & CMB-S4-like & PICO \\
\hline \hline
$C_\ell^{BB}$ & 1.0\footnote{This bound is based on fitting all cosmological parameters to TT, EE, ET from Planck and BB from SPT. The forecasts in the remained of the row assume fitting $B_{\rm SI}$ to BB only with remaining cosmological parameters fixed to their best fit LCDM values.} & 2.3  & 1.0   & 0.55 & 0.5  \\
${\rm FR}_{f_G=0}$ & - & 1.7  & 0.7   & 0.16 & 0.08  \\
${\rm FR}_{f_G=1}$ & 50 & 1.7  & 0.7   & 0.18 & 0.12  \\
 \hline
\end{tabular}
\caption{\label{tab:pmf} Current and forecasted $68$\% CL lower bounds on the strength of the scale-invariant primordial magnetic field $B_{\rm SI}$ derived from mode-coupling correlations induced by Faraday Rotation (FR) compared to those derived from the BB spectra ($C_\ell^{BB}$). The forecast accounts for de-lensing and beam systematics. The $f_G=0$ case assumes that there is no galactic contribution to FR, while $f_G=1$ includes the galactic FR based on the rotation measure map of  \cite{Oppermann:2011td}.
}
\end{table}

Table~\ref{tab:pmf} shows the 68\% CL bounds on the scale-invariant PMF expected from the FR measurements, and compare them to the bounds one would obtain by constraining the (non-FR) vector and tensor mode contributions of the PMF to the BB spectrum. As one can see, while the BB based constraints are stronger today, they will not significantly improve on the present $1$~nG bound. On the other hand, the FR based estimates will eventually do better, thanks for the linear scaling of the SNR with the PMF strenght. 

Importantly, experiments like CMB-S4 and PICO can achieve bounds on the PMF strength $\sim 0.1$~nG, which is a critical threshold for ruling out the purely primordial (no dynamo) origin of the $\sim 1-10$~$\mu$G galactic magnetic fields. Namely, a $0.1$~nG field coherent over a $1$ Mpc size region would be adiabatically compressed into a $\sim 1$~$\mu$G field in the galactic halo \cite{Grasso:2000wj}. 

The FR caused by a $\sim 0.1$~nG PMF is approximately the same as that due to the magnetic field in the Milky Way near the galactic poles \cite{De:2013dra}. Thus, lowering the FR based bound on the PMF below $0.1$ nG would require an independent measurement of the galactic RM. This should be possible in the future with improved versions of the galactic RM maps \cite{Oppermann:2011td} based on studies of extragalactic radio sources. Regardless of that, experiments like CMB-S4 and PICO will have the sensitivity to use FR to probe the magnetic field in our galaxy. Since FR probes the line-of-sight component of the magnetic field, it is complementary to studies using synchrotron radiation which probe the transverse component.

\subsection{Model-independent constraints on Lorentz-violating physics}
\label{sec:lvp}

The last two decades have seen a resurgence of interest in tests of Lorentz invariance due, in part,
to the suggestion that violations of Lorentz invariance could emerge in theories of quantum gravity \cite{Kostelecky:1988zi,Kostelecky:1991ak,Kostelecky:1994rn}. Of the hundreds of searches for Lorentz violation in particles and in gravity
\cite{Bluhm:2005uj,Tasson:2014dfa,Kostelecky:2008ts}, tests involving astrophysical sources are among the most sensitive since tiny Lorentz-violating defects can accumulate over long propagation times \cite{Ellis:2011ek}. CMB radiation is the oldest light available
to observation and provides extreme sensitivity to certain forms of Lorentz violation
\cite{Feng:2006dp,
  Cabella:2007br,
  Kostelecky:2007fx,
  Xia:2007qs,
  Xia:2008si,
  Kostelecky:2008be,
  Komatsu:2008hk,
  Wu:2008qb,
  Kahniashvili:2008va,
  Pagano:2009kj,
  Brown:2009uy,
  Xia:2009ah,
  Komatsu:2010fb,
  Xia:2012ck,
  Gruppuso:2011ci,
  Hinshaw:2012aka,
  Kaufman:2013vbd,
  Mei:2014iaa,
  Galaverni:2014gca,
  Zhao:2015mqa,
  Aghanim:2016fhp,
  Gruppuso:2015xza,
  Molinari:2016xsy}.

Tests of Lorentz symmetry are aided by a theoretical framework known as the Standard-Model Extension (SME), which aims at providing a general all-encompassing self-consistent description Lorentz violation in both the Standard Model of particle physics and General Relativity \cite{Colladay:1996iz,Colladay:1998fq,Kostelecky:2003fs}. The early work on the SME was largely motivated by suggestions that Lorentz invariance may be spontaneously broken in string theory \cite{Kostelecky:1988zi,Kostelecky:1991ak,Kostelecky:1994rn}. Other possible origins of Lorentz violation include small spacetime variations of physical ÒconstantsÓ or unconventional fields \cite{Kostelecky:2002ca,Bertolami:2003qs}, theories involving noncommutative spacetime \cite{Hayakawa:1999yt,Carroll:2001ws} and unconventional coupling to gravity \cite{Shore:2004sh}. In the SME, photons are described by the usual Maxwell Lagrangian augmented by an infinite series of Lorentz-violating terms
\cite{Kostelecky:2009zp},
\begin{align}
  {\cal L}_{\rm LV} &= 
  \tfrac12 \epsilon^{\kappa\lambda\mu\nu}
  A_\lambda (\kafd{3})_\kappa F_{\mu\nu}
  -\tfrac14 F_{\kappa\lambda} (\kfd{4})^{\kappa\lambda\mu\nu} F_{\mu\nu}
  \notag \\ &\quad
  + \tfrac12 \epsilon^{\kappa\lambda\mu\nu}
  A_\lambda {(\kafd{5})_\kappa}^{\alpha\beta}\partial_\alpha\partial_\beta F_{\mu\nu}
  + \ldots \ .
  \label{SME:lagrangian}
\end{align}
Each term in the series gives a different class of Lorentz violation controlled by tensor coefficients ${(\kafd{d})_\kappa}^{\alpha_1\ldots\alpha_{(d-3)}}$ and $(\kfd{d})^{\kappa\lambda\mu\nu\alpha_1\ldots\alpha_{(d-4)}}$. For example, the axion-photon coupling term in Eq.~(\ref{eq:pseudophi}) is physically equivalent to $\mathcal L_\text{CS} = - (\partial_\kappa a/2f_a) \epsilon^{\kappa\lambda\mu\nu} A_\lambda F_{\mu\nu}$, yielding a correspondence between the gradient of the axion field and the $d=3$ coefficients for Lorentz violation,
\begin{equation}
  (\kafd{3})_\kappa = - f^{-1}_a \partial_\kappa a \ .
  \label{SME:CS}
\end{equation}
The label $d=3,4,5,\ldots$ is the mass dimension of the conventional piece appearing with the coefficient, and it is expected that  lower-$d$ terms dominate at attainable energies. Consequently, most tests of Lorentz symmetry in photons have focused on the leading-order $d=3$ and $d=4$ violations.

Each term in the Lagrangian (\ref{SME:lagrangian}) leads to vacuum birefringence, which can be tested with extreme precision using polarimetry of astrophysical sources. For $d\geq 4$, the effects on the polarization of light grow with photon energy, and are best constrained using high-energy sources \cite{Kostelecky:2006ta}. However, the lowest-order $d=3$ term gives energy-independent birefringence, so the CMB provides the ideal source for this class of violations. Lorentz violation of the CS type was first bounded at the level of $10^{-42}\,{\rm GeV}$ three decades ago in a study of polarization in radio galaxies \cite{Carroll:1989vb}. Since Lorentz violation generally comes with violations of rotational symmetry, the effects of birefringence are typically direction dependent, and the full-sky CMB can test anisotropic birefringence more effectively than point sources.

The $d=3$ Lorentz violations cause a simple rotation in the linear polarization. Integrating from recombination to today, the CMB polarization rotates about the line of sight $\hat n$ by an angle
\cite{Kostelecky:2008be}
\begin{equation}
  \alpha(\hat n) \simeq
  -T \sum_{lm} Y_{lm}(\hat n) \kVdjm{3}{lm} \ ,
  \label{SME:alpha}
\end{equation}
where $T \simeq 3.8^\circ / 10^{-43}$ GeV is the time since recombination in units convenient for studies involving the SME.
For convenience, we have expanded the CPR rotation angle $\alpha(\hat n)$ in spherical harmonics. There are four non-zero spherical coefficients,
$\kVdjm{3}{00}$,
$\kVdjm{3}{11}$,
$\kVdjm{3}{10}$, and 
$\kVdjm{3}{1(-1)} = -\big(\kVdjm{3}{11}\big)^*$, which are linear combinations of the four $d=3$ tensor coefficients $(\kafd{3})_\kappa$ (note that in the case of the photo-axion coupling, this corresponds to assuming constant gradients in Eq.~(\ref{SME:CS})).
While anisotropic birefringence in the CMB has been considered by a number of researchers
\cite{Kamionkowski:2010ss,Li:2013vga,Zhao:2014yna,Li:2014oia,Li:2015vea,Ade:2015cao,Array:2017rlf},
relatively few constraints exist on the three $l=1$ coefficients describing the potential dipole anisotropy in the CPR angle (but see {\it e.g} the analysis of the 2003 BOOMERANG data in \cite{Kostelecky:2007fx}).
A dipole in the CPR angle was recently measured in the analysis of Planck data \cite{Contreras:2017sgi}, where it was expressed using the form $\alpha(\hat n) = A_1\, \hat n \cdot \hat N$, where $A_1$ is the maximum $\alpha$ and $\hat N$ is the direction at which maximum rotation occurs. 
Ref.~\cite{Contreras:2017sgi} reports an amplitude of $A_1 = 0.32^\circ\pm0.10^\circ\pm0.08^\circ$ and a direction $\hat N$ at galactic coordinates $l=295^\circ\pm22^\circ\pm5^\circ, b=17^\circ\pm 17^\circ\pm 16^\circ$. Neglecting the covariance between the parameters, the corresponding 1$\sigma$ bounds on anisotropic SME coefficients are
\begin{align}
  \kVdjm{3}{10} &= (0.09\pm0.06) \times 10^{-43}\, {\rm GeV} \ ,
  \notag \\
  {\rm Re}\,\kVdjm{3}{11} &= (-0.07\pm0.05) \times 10^{-43}\, {\rm GeV} \ ,
  \notag \\
  {\rm Im}\,\kVdjm{3}{11} &=  (0.00\pm0.04) \times 10^{-43}\, {\rm GeV} \ ,
\end{align}
representing an improvement of more than two orders of magnitude over previous bounds. 

\begin{table}[tbp]
\begin{tabular}{ccc}
   $\kVdjm{3}{00}$ ($10^{-43}\,{\rm GeV}$) & Data & Ref. \\
  \hline\hline
    $ -0.51\pm0.75\pm0.45 $ &  QUaD & \cite{Wu:2008qb} \\
  $ 0.34\pm1.16\pm1.40 $ &  WMAP9 & \cite{Hinshaw:2012aka} \\
  $ -0.29\pm0.05\pm0.26 $ &  Planck & \cite{Aghanim:2016fhp} \\
  \hline
  $\sim 0.017$ & LiteBIRD & \\
  $\sim 0.007$ & SO & \\
   $\sim 0.0025$ & CMB-S4-like & \\
   $\sim 0.0015$ & PICO & \\
  \hline
\end{tabular}
\caption{\label{SME:tabl}
Current CMB birefringence constraints on the isotropic $d=3$ coefficient for Lorentz violation $\kVdjm{3}{00}$ and projected sensitivities for the future experiments.}
\end{table}

As one can see from Table~\ref{tab:rotation}, the bounds on the quadrupole of anisotropic rotation will improve by a factor of 3 with LiteBIRD, a factor of 10 with SO, a factor of 30 with CMB-S4-like and a factor of 60 with PICO. Correspondingly, a comparable improvement is expected for the dipole contribution and the corresponding SME coefficients.

The remaining coefficient for Lorentz violation $\kVdjm{3}{00}$ yields a uniform rotation by angle 
\begin{equation}
  \alpha_{\rm iso} =
  -\frac{T}{\sqrt{4\pi}}\, \kVdjm{3}{00}
  \simeq -\frac{1.1^\circ}{10^{-43}\,{\rm GeV}}
  \, \kVdjm{3}{00},
  \label{SME:iso_alpha}
\end{equation}
thus, the bounds on uniform CPR from Table~\ref{tab:rotation} can be readily converted into bounds on $\kVdjm{3}{00}$. The current bound from Planck  \cite{Aghanim:2016fhp} limits $\kVdjm{3}{00}$ to a few $\times 10^{-44}\,{\rm GeV}$. The projected constraints from LiteBIRD, SO, CMB-S4-like and PICO are given in Table \ref{SME:tabl}. The sub-arcminute sensitivity expected in Stage IV experiments should yield sensitivities better than $10^{-46}\,{\rm GeV}$, representing at least a hundred-fold improvement in our ability to test Lorentz violation.

Overall, the future bounds on CPR will not just improve the CMB bounds on Lorentz violation, but will provide the best overall constraints on the $d=3$ CPT  and Lorentz violation in photons, improving on the original Carroll, Field and Jackiw result \cite{Carroll:1989vb} by four orders of magnitude!

\section{Conclusions}
\label{sec:summary}

Using the primordial universe as a probe of fundamental physics is not a new idea. Yet, until now, such measurements were beyond the reach of practical investigation. Now, upcoming and future CMB experiments will dramatically improve our ability to constrain cosmic polarization rotation, opening opportunities for probing both conventional aspects of fundamental physics, and so-called ``physics beyond the standard model". In particular, as we have shown, these results will significantly improve the bounds on  axion-photon coupling, coming close to excluding the entire class of string theory axions. More generally, they will put new stringent bounds on Lorentz violation in the universe, with important implications for model-building in high energy physics and quantum gravity. Crucially, these bounds are of a completely complementary nature to laboratory-based probes, which will add confidence if these exotic effects are ever discovered.

Experiments like CMB-S4 and PICO will improve bounds on primordial magnetic fields, achieving constraints close to the critical threshold of $0.1$ nG, which would rule out the purely primordial (i.e., no dynamo) origin of the observed $\mu$G level magnetic fields in galaxies. These observations will also open the possibility to use Faraday Rotation of CMB polarization as a probe the line-of-sight component of the magnetic field in our galaxy, complementing information obtained from the galactic synchrotron radiation that probes the transverse component of the field. 

\acknowledgments

We thank Vera Gluscevic and Eric Hivon for useful discussions and Colin Hill for providing the simulated PICO noise curves. This is not an official paper by the Simons Observatory or the CMB-S4 collaborations. L.P. is supported in part by the National Sciences and Engineering Research Council (NSERC) of Canada. The work of M.S. at Tel Aviv University has been supported by a grant from the Jewish Community Foundation (San Diego, CA). M.M. was supported in part by the United States National Science Foundation under grant number PHY-1819412. This research was enabled in part by support provided by WestGrid and Compute Canada.


%

\end{document}